# Jensen Inequalities for Tunneling Probabilities in Complex Systems

D. M. Andrade and M. S. Hussein

*Departamento de Física Matemática,*
*Instituto de Física, Universidade de São Paulo*
*C.P. 66318, 05314-970 São Paulo, S.P., Brazil*

The Jensen theorem is used to derive inequalities for semi-classical tunneling probabilities for systems involving several degrees of freedom. These Jensen inequalities are used to discuss several aspects of sub-barrier heavy-ion fusion reactions. The inequality hinges on general convexity properties of the tunneling coefficient calculated with the classical action in the classically forbidden region.

In complex quantum systems with several active degrees of freedom, one usually finds a strong deviation of the tunneling probability from the prediction of simple one-dimensional barrier penetration model. Experiments on the fusion of nuclei at sub-barrier energies have clearly shown a very large enhancement of the tunneling probability when compared to simple one-dimensional barrier model calculations. In several important theoretical papers [1], [2], [3], [4], [5] addressing the tunneling problem in systems coupled to a reservoir attempts were made to obtain semi-quantitative, albeit important estimates of the effects of the reservoir's degrees of freedom on the tunneling dynamics of the subsystem of interest. More detailed numerical calculations based on the coupled-channels description of e.g. sub-barrier fusion attempt to give quantitative description within a restricted dimension of the reservoir (the number of channels strongly coupled to the entrance channel) [6], [7], [8], [9].

Furthermore, very low energy fusion of light nuclei such as $^2$H + $^2$H has been of interest over the last two decades in the context of the so-called cold fusion. In this endeavour the effect of the environment is important. Recent work on the fusion of such light nuclei has indicated that in metals electron screening is enhanced, reducing the fusion barrier and accordingly enhancing fusion probability. Of course such light ion reactions are of great importance in astrophysics [10] and the understanding of the effect of the environment on them has been under intensive experimental [11], [12], [13] and theoretical [14], [15], [16], [17] scrutiny. It would be a useful compliment to the above discussion to find general inequalities that involve the tunneling probability for a sub-system of the many-degrees-of-freedom system when compared to the sub-system alone (with the coupling to the reservoir being averaged). This is the aim of the present work. We rely on a general theorem in analysis referred to as the Jensen theorem.

The Jensen inequality [18] ensures that if $F(f(\zeta))$ is a functional of a function $f(\zeta)$, then $\langle F(f(\zeta))\rangle_\zeta \geqslant F\left(\langle f(\zeta)\rangle_\zeta\right)$ if and only if $F$ is a convex functional of $f$ within the interval in which the average $\langle\ \rangle_\zeta$ is being calculated. One possible way of explicitly stating the Jensen inequality is the following:

$$\frac{\int_a^b d\zeta \phi(\zeta) F(f(\zeta))}{\int_a^b d\zeta \phi(\zeta)} \geq F\left[\frac{\int_a^b d\zeta \phi(\zeta) f(\zeta)}{\int_a^b d\zeta \phi(\zeta)}\right] \tag{1}$$

if and only if $F(f)$ is a convex functional of $f$ within the interval [a, b], and $\phi(\zeta)$ is any positive integrable function. If the convexity turns out to be a concavity, the inequality is reversed.

One immediate consequence of the Jensen inequality is Peierls theorem, which was used by Peierls [20] to prove that the canonical partition function, $Z(\beta)$, defined by $Z(\beta) = Tr[\exp[-\beta H]]$, is greater or equal than $\exp[-\beta Tr H]$.

Using Peierls theorem, R. Johnson and C. Goebel (JG) derived an inequality involving the reflection above the barrier in order to assess the effect of breakup on the elastic scattering of halo nuclei [19]. That inequality clarified why the reaction cross section calculated within the Glauber model is appreciably smaller than that calculated using the optical limit of the model, a point first emphasized in [21], thus resulting in larger radii of halo nuclei. In the following we show that the result of JG [19] and that of [21] can be considered as a consequence of the Jensen inequality.

In their above cited work, JG considered the elastic S-matrix element for the $l^{th}$ partial wave

$$S_l(E, \zeta) = \exp[2i\delta_l(E, \zeta)] = \exp[f] \tag{2}$$

where the phase shift $\delta_l$, in the JWKB approximation is given by,

$$\delta_l(E, \zeta) = \lim_{r \to \infty} \left\{ \int_{r_0}^r dr'(k_l(r', \zeta) - \int_{r_0^{(0)}}^r dr' k_l^{(0)}(r')) \right\} \tag{3}$$

Above, $k_l(r, \zeta)$ is the local wave number given by $k_l(r, \zeta) = \sqrt{\frac{2\mu}{\hbar^2}[E - V_l(r) - F(r)G(\zeta)]}$, $k_l^{(0)}(r)$ is the free particle local wave number, $r_0$ is the classical turning point defined by $k_l(r_0, \zeta) = 0$ and $r_0^{(0)}$ is the corresponding one for the free local wave number. The asymptotic wave number is denoted by $k = k_l^{(0)}(r = \infty)$ and the mass by $\mu$.

At high energies, one may expand the local wave number in powers of $\frac{V(r) + F(r)G(\zeta)}{E}$ and retain the leading term. This constitutes the Eikonal approximation considered by JG [19]. This approximation gives for the phase shift,

$$\delta_{Eikonal}(E, b, \zeta) = -\frac{\mu}{\hbar^2 k} \int_b^\infty r dr \frac{V(r) + F(r)G(\zeta)}{\sqrt{r^2 - b^2}} \tag{4}$$

where the impact parameter $b = \frac{l+1/2}{k}$. We consider, as JG, the case where the potential, and accordingly the form factor, is purely absorptive ($V(r) = -iW(r)$ and $F(r) = \frac{V(r)}{dr}$). Then the phase shift $\delta_{Eikonal}$ becomes pure imaginary and $f$ real. From the Jensen inequality and from the fact that $\delta_{Eikonal}(E, b, \zeta)$ is a linear function of $G(\zeta)$ we obtain the following inequality

$$\overline{S_l(E, \zeta)} \geq \exp[-2|\overline{\delta_{Eikonal}(E, b, \zeta)}|] \tag{5}$$

which is the result obtained in the work of JG.

The above inequality only holds for imaginary phase shifts. Clearly the actual heavy-ion scattering at intermediate energies involves complex phase shifts, and this fact points to an inherent limitation of the work of JG. This limitation is removed if we go to very low energies and consider fusion which is dominated by quantum tunneling (with real action integral).

In order to apply the Jensen inequality to fusion reactions, we recall first the expression for tunneling probability provided by the coupled-channel treatment in the case of a coupling to an oscillator reservoir with zero frequency (sudden approximation), which can be cast as a simple average [22]:

$$\langle T_l(E) \rangle_\zeta \equiv \int d\zeta |\phi_0(\zeta)|^2 T_l \left[ E, V_l(r) + H_{int}(r, \zeta) \right] \tag{6}$$

where $T_l [E, V_l(r) + H_{int}(r, \zeta)]$ is the transmission probability evaluated at energy $E$ with an effective potential $V_l(r) + H_{int}(r, \zeta)$, and the wave function, $\phi_0(\zeta)$, denotes the ground state wave function related to the reservoir coupling. Of course wave functions for excited states of the considered reservoir coupling can be used instead. The above equation refers to the limit in which the intrinsic energies are small compared to the coupling interaction, so that the reservoir Hamiltonian is set equal to zero.

Using the Kemble [23] form of the transmission probability [25], which guarantees a 1/2 transmission at the top of a symmetrical barrier [24], and takes into account multiple reflections inside the barrier to all order if the uniform approximation is used in a path integral formulation of tunneling [25], $T_l [E, V_l(r) + H_{int}(r, \zeta)]$ is found to be

$$T_l [E, V_l(r) + F(r) G(\zeta)] = \frac{1}{1 + exp\{g_l [E, V_l(r) + H_{int}(r, \zeta)]\}}, \tag{7}$$

with $g_l [E, V_l(r) + H_{int}(r, \zeta)]$ given by

$$g_l [E, V_l(r) + F(r) G(\zeta)] = \sqrt{\frac{8\mu}{\hbar^2}} \int_{r_1(l,\zeta)}^{r_2(l,\zeta)} dr \sqrt{V_l(r) + H_{int}(r, \zeta) - E} \tag{8}$$

where $r_1(l, \zeta)$ and $r_2(l, \zeta)$ are the classical turning points. Here we consider only the case where the form factor $H_{int}(r, \zeta)$ is fixed at the position of the maximum of the barrier $V_l(r)$, namely $H_{int}(r, \zeta) = H_{int}(R_l, \zeta)$. Although this is a very rough approximation, it is a first step in the direction of assessing the effects of the contribution of the coupled potential on the transmission coefficient. Bringing the Jensen inequality into the context of fusion probability, one can state that

$$\langle T_l(E) \rangle_\zeta \geq T_l \left[ E, V_l(r) + \langle H_{int}(R_l, \zeta) \rangle_\zeta \right] \tag{9}$$

if and only if $T[E, V_l(r) + H_{int}(R_l, \zeta)]$ is a convex functional of $H_{int}(R_l, \zeta)$. In the equation above, $\langle H_{int}(R_l, \zeta) \rangle_\zeta$ is defined as $\langle H_{int}(R_l, \zeta) \rangle_\zeta \equiv \int_a^b d\zeta |\phi_0(\zeta)|^2 H_{int}(R_l, \zeta)$ and $|\phi_0(\zeta)|^2$ is the square modulus of the normalized ground-state wave function related to the reservoir. The same definition holds for the average $\langle T_l(E) \rangle_\zeta$. Hence, it is necessary to determine whether the transmission probability is a convex or a concave function of $H_{int}(R_l, \zeta)$ (where $H_{int}(R_l, \zeta)$ is regarded as a simple variable) in order to make a comparison of the type of inequality (9). The interval $[a, b]$ stands for all possible values that the coordinate related to the oscillator reservoir, $\zeta$, may assume.

Let us introduce the quantity $w(\zeta) = E - H_{int}(R_l, \zeta)$, which will be used in our calculations in order to make the physical comprehension clearer, that is, $w(\zeta)$ will stand for the effective energy. Because $w(\zeta)$ is a linear function of $H_{int}(R_l, \zeta)$, the sign of the second derivative of the tunneling probability $T_l$, with respect to $w(\zeta)$ determines if $T_l$ is a convex functional of the function $H_{int}(R_l, \zeta)$ or a concave one:

$$\frac{\partial^2 T_l}{\partial w^2} = \frac{\exp[h_l(w)]}{(1+\exp[h_l(w)])^3} \left\{ (\exp[h_l(w)] - 1)(f_l(w))^2 + (\exp[h_l(w)] + 1) \left( \frac{\partial f_l(w)}{\partial w} \right) \right\} \tag{10}$$

in which $h_l(w) = \sqrt{\frac{8\mu}{\hbar^2}} \int_{r_1(l,w)}^{r_2(l,w)} dr \sqrt{V_l(r) - w}$ and $f_l(w) = \sqrt{\frac{2\mu}{\hbar^2}} \int_{r_1(l,w)}^{r_2(l,w)} \frac{dr}{\sqrt{V_l(r) - w}}$.

For heavy ions at near-barrier energies, the effective tunneling potential $V_l(r)$ is usually approximated by a an inverted parabola, which enables us to follow the Hill-Wheeler procedure [26], in order to obtain a closed form for the Kemble tunneling probability. For such heavy ions, the extra degree of freedom, namely the coordinate $\zeta$, would stand for the displacement due to vibrational modes, and the coupled reservoir would be represented by an oscillator in this case. Therefore, for such cases,

$$V_l(r) = V_{HWl}(r) \equiv V_l(R_l) - \frac{1}{2}\mu\omega_l^2 (r - R_l)^2 \tag{11}$$

Hence,

$$f_l(w) = \sqrt{\frac{2\mu}{\hbar^2}} \int_{r_1(l,w)}^{r_2(l,w)} \frac{dr}{\sqrt{V_l(R_l) - \frac{1}{2}\mu\omega_l^2 (r - R_l)^2 - w}} = \frac{2\pi}{\hbar\omega_l} \Rightarrow \frac{\partial f_l(w)}{\partial w} = 0 \tag{12}$$

This result combined with Eq. (10) yields:

$$\frac{\partial^2 T_l}{\partial w^2} = \frac{\exp[h_l(w)]}{(1+\exp[h_l(w)])^3} (\exp[h_l(w)] - 1)(f_l(w))^2 > 0 \tag{13}$$

and finally,

$$\langle T_l[w(\zeta), V_{HWl}(r)] \rangle_\zeta \geq T_l \left[ \langle w(\zeta) \rangle_\zeta, V_{HWl}(r) \right]$$

or, since $\langle w(\zeta) \rangle_\zeta = E - \langle H_{int}(R_l, \zeta) \rangle_\zeta$,

$$\langle T_l[E, V_{HWl}(r) + H_{int}(R_l, \zeta)] \rangle_\zeta \geq T_l \left[ E, V_{HWl}(r) + \langle H_{int}(R_l, \zeta) \rangle_\zeta \right] \tag{14}$$

for all $l$-partial wave functions and different values of $\mu$. Thus within Hill-Wheeler approximation, all systems show enhanced tunneling.

We now assess the application of the Jensen inequality for the case of low energies. By "*low*" energies, we mean small values of the function $w(\zeta)$. Here the coupling to the reservoir could stand for coupling to the electronic degrees of freedom. The parabolic approximation for the potential barrier is not suitable for this case, and we shall use a general ion-ion effective interaction, which has the form

$$V_l(r) \equiv V_N(r) + \frac{Z_1 Z_2 e^2}{r} + \frac{\hbar^2 l(l+1)}{2\mu r^2} \qquad (15)$$

where $V_N(r)$ is the nuclear attractive potential. As shown in the Appendix, whatever specific analytic form the short-range nuclear interaction, $V_N(r)$, may take, a potential barrier resulting from $V_l(r)$ of Eq.(16) always leads to the following important Jensen inequality for very small values of $w$ and/or $E$:

$$\langle T_l\left[E, V_l(r) + H_{int}(R_l, \zeta)\right]\rangle_\zeta \geq T_l\left[E, V_l(r) + \langle H_{int}(R_l, \zeta)\rangle_\zeta\right] \qquad (16)$$

where $V_l(r)$ is defined by Eq. (15).

This very general result implies that whatever attractive nuclear potential model one may use, the plot of the curve of the transmission probability versus $w(\zeta)$, where $w(\zeta) = E - H_{int}(R_l, \zeta)$, is always convex for small values of $w$, leading to enhanced tunneling.

From the results depicted by Eqs. (15) and by Eq. (17), one is compelled to infer that, in general, the tunneling probability $T_l$ would tend to be a convex functional of the function $w(\zeta)$, as seen in Fig.1, in which the tunneling probability was defined by Eq. (7), and $w(\zeta)$ ranges from $E_{\min}$ to $E_{\max}$, where $E_{\min}$ is the minimal energy required for the transmission probability to be finite and $E_{\max}$ stands for the hight of the tunneling barrier, for each considered system. This general property of convexity implies an enhanced tunneling or fusion, as experimental data seem to clearly indicate [27].

So far we have concentrated our attention on the $l^{th}$ transmission coefficient. The experimental data, on the other hand, are represented by the fusion cross section defined by,

$$\sigma_F(E) = \frac{\pi \hbar^2}{2\mu E} \sum_{l=0}^{\infty} (2l+1) T_l(E) = \sum_{l=0}^{\infty} \sigma_l(E) \qquad (17)$$

From Eq. (17), we see that the dependence of $\sigma_F(E)$ on the coupling $H_{int}(R_l, \zeta)$ lies only on the terms $T_l(E)$. Then, if it is possible to state that, for instance, $T_l(E)$ is a convex functional of $H_{int}(R_l, \zeta)$ for all values of the quantum number $l$, then one can also state that $\sigma_F(E)$ is a convex functional of $H_{int}(R_l, \zeta)$. The above lends support to the general idea that there is an enhancement of the fusion cross section when coupling to the degrees of freedom of the reservoir are taken into account, namely

$$\langle \sigma_F(w(\zeta))\rangle_\zeta \geq \sigma_F(\langle w(\zeta)\rangle_\zeta) \qquad (18)$$

This is easily seen at deep sub-barrier energies, where in fact the transmission coefficient or tunneling probability can be approximated by an exponential, since the action in the Kemble formula is small,

$$\sigma_F(E) = \frac{\pi \hbar^2}{2\mu E} T_0(E) = \frac{\pi \hbar^2}{2\mu E} \exp\left[-g_0(E, V(r) + H_{int}(R_l, \zeta))\right] \qquad (19)$$

As shown in the Appendix, the above function is convex in $H_{int}(R_l, \zeta)$ for $w(\zeta) \to 0$, and thus its average over $\zeta$ is greater than that calculated with $\langle H_{int}(R_l, \zeta)\rangle_\zeta$. Thus, we can state,

$$\langle \exp\left[-g_0(E, V(r) + H_{int}(R_l, \zeta))\right]\rangle_\zeta \geq \exp\left[-g_0(E, V(r) + \langle H_{int}(R_l, \zeta)\rangle_\zeta)\right] \quad (20)$$

which represents the very low energy tunneling version of the JG inequality of elastic scattering eikonal S-matrix element of halo nuclei.

In reality, large enhancement in $\sigma_F$ has been observed for most heavy-ion fusion systems at sub-barrier energies [27]. Recently, it has been reported that at deep sub-barrier energies, this enhancement is reduced [28] (unfortunately, this effect has been widely called hindrance, which should not be confused with what we mean by hindrance, namely, a cancave behavior of $T_l(w(\zeta))$ as a function of $H_{int}(R_l, \zeta)$).

The convexity of the unaveraged tunneling probability for $l = 0$ can be seen in Fig. 1 for the systems $^{64}Ni+^{64}Ni$ and $^{16}O+^{150}Sm$. However, plots of the tunneling probability as a function of $w(\zeta)$ for very light ions ($^2H+^2H$, $^3H+^3H$), as shown in Fig. 2, show a different behavior. For such light ions, the curve of $T_l$ versus $w(\zeta)$ presents an inflection point, and thus for high values of $w$ it becomes concave. This result is in contradiction with the general analytical result represented by Eq. (14), where the parabolic potential was used to approximate the real potential barrier. That happens possibly because for light ions such approximation for the potential barrier is not suitable, as it appears that a more accurate approximation that would take into account the highly asymmetrical character of the potential curve would be required. A third degree polynomial would be a better fit for this purpose, but the analytical treatment becomes extraordinarily more complicated. For fusion probabilities involving light ions, the Jensen inequality can only be applied within restricted regions of the spectrum of values which $w(\zeta)$ may assume.

In conclusion we have considered some general properties of the tunneling probability for systems coupled to a reservoir. Using the Jensen inequality, we have shown that within the Kemble/uniform approximation theory of the tunneling probability, the average transmission probability is in general larger than that calculated when the reservoir degrees of freedom are averaged out at the outset. This has an immediate consequence on sub-barrier fusion of heavy ions, where data seem to indicate an enhanced tunneling owing to the coupling to the reservoir (coupled channels effects). In addition, we have shown that the results obtained by JG [19] can be generalized by using the Jensen inequality. The underlying mathematical dependence of the tunneling probability as a function of the reservoir coupling, namely the tunneling probability is in general a convex functional of the coupling hamiltonian, permits Jensen inequality to be applied to this research field in order to compare two different forms of reaction probabilities, both of physical interest. The peculiar behavior presented by the curves of transmission probability for light ions indicates that for such systems the effect of the coupling to the reservoir on fusion might be different from such effect in heavier nuclei, since a change to concavity reverses the Jensen inequality, and the enhancement in tunneling becomes a hindrance.

The authors thank Professor João Barata for very instructive discussions. This work was supported in part by the Brazilian agencies, CNPq and FAPESP. MSH was the 2007/2008 Martin Gutzwiller Fellow at the Max-Planck-Institute for the Physics of Complex Systems (MPIPKS) in Dresden, where part of this work was carried out. Both authors thank the MPIPKS-Dresden for hospitality and support.

**APPENDIX**

In this appendix we apply the Jensen inequality for the tunneling probability for very small $w$ and/or $E$, Eq. (17).

From Eq. (10), it follows that for small values of $w(\zeta)$, one has

$$\frac{\partial^2 T_l}{\partial w^2} \approx \frac{\exp[2h_l(w)]}{(1+\exp[h_l(w)])^3}\left\{(f_l(w))^2 + \left(\frac{\partial f_l(w)}{\partial w}\right)\right\} \quad (21)$$

where $f_l(w)$ and $h_l(w)$ are defined as in Eq. (10). From the equation above, we see that the sign of $\frac{\partial^2 T_l}{\partial w^2}$ will depend exclusively on the term $\left\{(f_l(w))^2 + \left(\frac{\partial f_l(w)}{\partial w}\right)\right\}$. We will show that such term, considering the potential barrier for fusion reaction with which we are dealing (Eq. (15)), is always positive when $w(\zeta)$ tends to zero. In order to do this, we first assume the contrary, namely we suppose that $\left\{(f_l(w))^2 + \left(\frac{\partial f_l(w)}{\partial w}\right)\right\}_{w\to 0} \leqslant 0$. Then,

$$\lim_{w\to 0}\left\{-\frac{d}{dw}\left(\frac{1}{f_l(w)}\right)\right\} \leqslant -1 \Rightarrow 1 \geqslant \lim_{w\to 0}\{wf_l(w)\} \Rightarrow 1 \geqslant \lim_{w\to 0}\left\{\sqrt{\frac{2\mu}{\hbar^2}}w\int_{r_1(l,w)}^{r_2(l,w)}\frac{dr}{\sqrt{V_l(r)-w}}\right\}$$

Now, let us make

$$\lim_{w\to 0}\left\{\int_{r_1(l,w)}^{r_2(l,w)}\frac{dr}{\sqrt{V_l(r)-w}}\right\} = \lim_{w\to 0}\left\{\int_{r_1(l,w)}^{r^*}\frac{dr}{\sqrt{V_l(r)-w}}\right\} + \lim_{w\to 0}\left\{\int_{r^*}^{r_2(l,w)}\frac{dr}{\sqrt{V_l(r)-w}}\right\}$$

in which $r_1(l,w) < r^* < r_2(l,w)$. Here $r^*$ is chosen to be greater than the distance at which the attractive nuclear potential becomes negligible. Hence, for $w \to 0$, we have

$$1 \geqslant \lim_{w\to 0}\left\{\sqrt{\frac{2\mu}{\hbar^2}}wI_1\right\} + \lim_{w\to 0}\left\{\sqrt{\frac{2\mu}{\hbar^2}}wI_2\right\} \quad (22)$$

where $I_1 \equiv \int_{r_1(l,w)}^{r^*}\frac{dr}{\sqrt{V_l(r)-w}}$ and $I_2 \equiv \int_{r^*}^{r_2(l,w)}\frac{dr}{\sqrt{V_l(r)-w}}$. Clearly $I_1$ is bounded for all values of $w \to 0$, and therefore $\lim_{w\to 0}\left\{\sqrt{\frac{2\mu}{\hbar^2}}wI_1\right\} = 0$. That leave us with the inequality

$$1 \geqslant \lim_{w\to 0}\left\{\sqrt{\frac{2\mu}{\hbar^2}}wI_2\right\} \quad (23)$$

We now turn to the question whether $wI_2$ is bounded for $w \to 0$. Performing a change of variables, namely $y = V_l(r) - w$, one gets for $I_2$:

$$I_2 = \int_{V(r^*)-w}^{0}\frac{dy}{\sqrt{y}}\frac{dV_l^{-1}(y+w)}{dy}$$

Since the point $r^*$ is taken to be much greater than the effective nucleus radius, the contribution for the total potential $V_l(r)$ of the attractive Woods-Saxon potential can be neglected within the interval $(r^*, r_2(l,w))$. Therefore, in the calculations for $I_2$, we approximate

$$V_l(r) \approx \frac{C_1}{r} + \frac{C_{2l}}{r^2} \qquad (24)$$

in which $C_1 = Z_1 Z_2 e^2$ and $C_{2l} = \frac{\hbar^2 l(l+1)}{2\mu}$. Clearly $C_1$ and $C_{2l}$ are non-negative. Here we first assume that $l \neq 0$, and therefore $C_{2l}$ is strictly positive. From Eq. (24), we have

$$r = \frac{C_1 + \sqrt{C_1^2 + 4(y+w)C_{2l}}}{2(y+w)}$$

and accordingly

$$I_2 = \int_{\frac{C_1}{r^*} + \frac{C_{2l}}{r^{*2}} - w}^{0} dy \left[ \frac{C_{2l}}{(y+w)\sqrt{y(C_1^2 + 4(y+w)C_{2l})}} - \frac{C_1 + \sqrt{C_1^2 + 4(y+w)C_{2l}}}{2\sqrt{y}(y+w)^2} \right]$$

It is not difficult to prove that $\frac{\partial I_2}{\partial C_1} > 0$. A direct consequence of this fact is that $\lim_{C_1 \to 0} \{I_2(C_1)\} \leqslant I_2(C_1)$, since $C_1$ is positive. Hence

$$\frac{\sqrt{C_{2l}}}{w}\sqrt{1 - \frac{w(r^*)^2}{C_{2l}}} \leqslant I_2 \Rightarrow \sqrt{C_{2l}} \leqslant \lim_{w \to 0} \{wI_2(w)\}$$

Combining the last result with the inequation (23) we find

$$1 \geqslant \sqrt{\frac{2\mu}{\hbar^2}} \lim_{w \to 0} \{wI_2(w)\} \geqslant \sqrt{\frac{2\mu}{\hbar^2} C_{2l}}$$

which implies the absurd result that $\sqrt{l(l+1)} \leq 1$ since $C_{2l} = \frac{\hbar^2 l(l+1)}{2\mu}$. By assumption, $l \neq 0$, and hence the minimum value for the term $\sqrt{l(l+1)}$ is $\sqrt{2}$. That leads us to a contradiction, and therefore our initial assumption, $\left\{(f_l(w))^2 + \left(\frac{\partial f_l(w)}{\partial w}\right)\right\}_{w \to 0} \leqslant 0$ can not be true. Let us examine now the case where $l = 0$, which means that the effetive potential $V_l(r)$ used in $I_2$ will be just the Coulomb potential:

$$V_0(r) \approx \frac{C_1}{r}$$

With the above potential, the integral $I_2$ becomes, for $w \to 0$:

$$I_2 = \frac{r^*}{\sqrt{w}}\sqrt{\frac{C_1}{wr^*} - 1} + \frac{2C_1}{w^{\frac{3}{2}}} \arctan\left\{\exp\left[\operatorname{arccosh}\left(\sqrt{\frac{C_1}{wr^*}}\right)\right]\right\} - \frac{\pi C_1}{2w^{\frac{3}{2}}} \qquad (25)$$

Multiplying both sides of Eq. (25) by $w$ and taking the limit $w \to 0$, we find that $\lim_{w \to 0} \{wI_2(w)\} = \infty$, and therefore the inequality (23) can neither be satisfied for the case of the partial wave with $l = 0$, nor the case $l \neq 0$. This proves that the assumption we made at the beginning of this section, namely $\left\{(f_l(w))^2 + \left(\frac{\partial f_l(w)}{\partial w}\right)\right\}_{w \to 0} \leqslant 0$, is false. Therefore, recalling Eq. (21), we have that for small values of $w(\zeta)$, $\frac{\partial^2 T_l}{\partial w^2} > 0$, which implies, for $w \to 0$, that

$$\langle T_l \left[E, V_l(r) + H_{int}(R_l, \zeta)\right]\rangle_\zeta \geq T_l \left[E, V_l(r) + \langle H_{int}(R_l, \zeta)\rangle_\zeta\right] \tag{26}$$

where $V_l(r)$ is defined by Eq. (15).

---

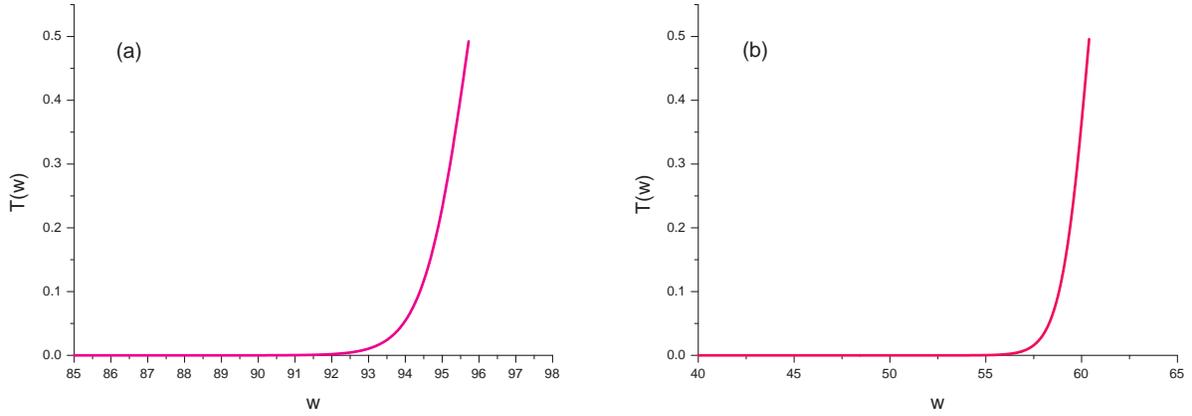

FIG. 1: Tunneling probability for $l = 0$ versus the function $w(\zeta)$ for the systems a) $^{64}$Ni + $^{64}$Ni, and b) $^{16}$O + $^{150}$Sm. The curves show a convex dependence of the tunneling probability functional on the function $w(\zeta)$.

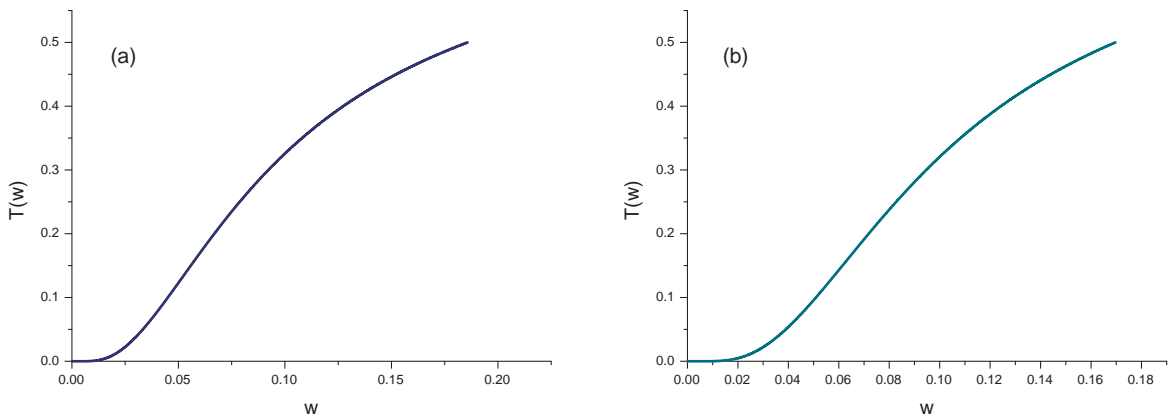

FIG. 2: Tunneling probability for $l = 0$ versus the function $w(\zeta)$, for the systems a) $^{2}$H+$^{2}$H, and b) $^{3}$H+$^{3}$H. Both curves show a change in curvature of the tunneling probability functional as the function $w(\zeta)$ increases.